Su dongcai

# Compressed sensing with corrupted observations[1]

Dongcai Su at 2015/9/9

## Abstract

We proposed a weighted $\ell_1$ minimization: $\min_{x,f} \|x\|_1 + \lambda\|f\|_1 \text{ s.t. } Ax + f = b$ to recover a sparse vector $x^*$ and the corrupted noise vector $f^*$ from a linear measurement $b = Ax^* + f^*$ when the sensing matrix $A$ is an $m \times n$ row i.i.d subgaussian matrix. Our first result shows that the recovery is possible when the fraction of corrupted noise is smaller than a positive constant, provided that $\|x^*\|_0 \leq O(n/\ln(n/\|x^*\|_0))$, which is also the asymptotically optimal bound. While our second result shows that the recovery is still possible when the fraction of corruption noise grows arbitrary close to 1, as long as $m \geq O\|x^*\|_0 \ln(\|x^*\|_0)$, which is asymptotically better than the bound $m \geq O\|x^*\|_0 \ln(n) \ln(m)$ achieved by a recent literature [1] by a $\ln(n)$ factor.

## 1. Introduction

Compressed Sensing (CS) have been well-studied recently [2] [3, 4]and have achieved great successes in industrial applications such as single pixel camera [5], Magnetic Resonance Image (MRI) [6-9], Radar [10] etc. In traditional compressed sensing, the task is to recover signal $x^*$ from a collection of network data $b = Ax$, where $A \in R^{m \times n}$ is called the sensing matrix and $b \in R^m$ is the measurement vector or observation vector. Based on the analysis of Gelfand's widths of $\ell_1$-balls, given any $A$ and $b$, it can be shown that the minimal number of measurements m required for recovery of $x^*$ from b via any method is $m \geq O(\|x^*\|_0 \ln(n/\|x^*\|_0))$, Chp 10 of [11].

A well-studied algorithm to recovery $x^*$ from the measurement vector b is through the solution $\hat{x}$ of the below $\ell_1$ minimization:
$\min_x \|x\|_1 \text{ s.t. } Ax = b$        (1.1)

There are typically 2 types of probabilistic recovery guarantee for (1.1), the first type of guarantee is called the uniformly recovery guarantee chp 12 of [11], which states that the solution of (1.1) can recover $x^*$ with high probability for all $x^*$ whose cardinality is small enough. Whereas the second type of guarantee is called the non-uniformly recovery guarantee chp 12 of [11], which asserts the solution of (1.1) can recover $x^*$ with high probability for any fixed $x^*$.

---
[1] Part of this paper had been submitted to *The **2016** IEEE **International Symposium on Information Theory (isit 2016)***

Uniformly recovery guarantee for (1.1) can be achieved if the sensing matrix A satisfies the restricted isometry property (RIP) [12, 13]. it's now proved that a broad families of random matrices satisfy the RIP with high probability, e.g., if A is sub-gaussian matrix [14] chap.5 of [15], then A satisfies the RIP-2k condition if m$\geq$ O(kln (en/k)), where $k = \max\{1, \|x^*\|_0\}$ is a positive constant. Similarly, If A is a partial Bounded Orthogonal System (BOS)[14] chap.12 of [11], then A satisfies the RIP-2k condition provided that m$\geq$ O(kln$^4$ (n)). According to [16], the solution of (1.1) can recover $x^*$ for all $x^*$ which is sparse enough to ensure the RIP-2k condition of A.

The non-uniformly recovery guarantee for (1.1) can be obtained by constructing a dual vector which satisfies the dual certification theorem 4.30 of [11], typical theoretical results in this case are: 1) if A is sub-guassian matrix and m$\geq$ O(kln (en/k)), then the solution of (1.1) can recovery $x^*$ for any fixed $x^*$ chap. 9 of [11]. Similarly, if A is BOS, then non-uniformly recovery guarantee of (1.1) can be derived provide that m$\geq$ O(kln (en)) chap. 12 of [11].

In a more general case where the measurement noise contains some bounded measurement noise, e.g., $b = Ax^* + v$, where $v$ is a m-dimensional vector denoting the measurement noise whose norm is bounded by a constant $\eta \geq 0$: $\|v\|_2 \leq \eta$, in such case, $x^*$ is recovered through the solution of the below convex program:

$\min_x \|x\|_1$ s. t. $\|Ax - b\|_2 \leq \eta$ (1.2)

## 1.1 motivated applications

It can be shown that stable recovery [2]guarantee for (1.2) can also be established based on the RIP condition [16] or inexact duality conditions theorem 4.33 of [11]. However, if a few entries of b are grossly corrupted which cause $\eta$ to be extremely large, consequently, the solution of (1.2) may depart unexpectedly far away from $x^*$. Unfortunately, corruptions and irrelevant measurements can take place frequently in modern applications during data acquisition, data transmission due to the factors like devices flaws and environmental hazards [1, 17, 18]. We should just name a few representative applications lie in this case:

- Network data pollution. In a sensor network, sensors correct measurements of the same signal x independently, e.g., the i[th] sensor collects data $z_i = \langle a_i, x \rangle$, where $a_i$ is the i[th] row of the sensing matrix and then send the data back for analysis. Typical application examples in this case we refer to MRI [6-9] and radar imaging [10]. However through the process of data acquisition and transmission, data missing or nonlinear mapping can occur due to the hazard environment conditions, hardware failure etc, which lenders some data in b report totally irrelevant measurements.
- Object recognition. In the face recognition problem, Wright [18] treats $x^*$ as the measurement vector of the face, the corrupted noise includes unwanted objects such as glasses, scarf, hats etc which typically occupy a positive fraction of b.

---

[2] The stable recovery stated here means that the recovery error of (1.2) $\|x^* - \hat{x}\|_2$ is proportional to $\eta$.

Other applications such as subspace clustering [19], image inpainting [20] and joint source channels coding [21] can also experience certain amount of corrupted noise.

This motivates another line of works in CS, which is called compressed sensing with corruption [22, 23], where the measurement vector is represented as $b = Ax^* + f^*$, here $f^*$ is a m-dimensional vector denoting the corrupted noise, which is often assumed to be sparse but whose non-zero entries can take arbitrary values. Typically $x^*$ and $f^*$ are recovered through the solution $\hat{x}$ and $\hat{f}$ of below weighted $\ell_1$ minimization

$$\min_{x,e} \|x\|_1 + \lambda \|f\|_1 \text{ s.t. } Ax + f = b \quad (1.3)$$

## 1.2 Previous works

(1.3) is applied to separate the sinus and spikes when setting $\lambda = 1$ and A is Full Fourier basis in literatures [24, 25], it requires both $\|x^{(0)}\|_0$ and $\|f^{(0)}\|_0$ are bounded from above by $O(\sqrt{n})$ in order to achieve deterministic recovery guarantee. The upper-bound for $\|x^{(0)}\|_0$ and $\|f^{(0)}\|_0$ is relaxed to be $O(n/\sqrt{\ln(n)})$ in [26], however the recovery guarantee achieved in [26] is probabilistic and it impose extra random assumptions on $x^{(0)}$ and $f^{(0)}$. Later, Wright et.al [27] analyze model (1.3) motivated by the face recognition problem, they show that when A is a i.i.d Gaussian designed matrix, then the exact recovery of (1.3) is possible even when grows arbitrarily close to m, provided that $\|x^{(0)}\|_0$ is sub-linear smaller than m. A bit later [28] and [29] independently show that recovery of (1.3) with $\lambda = 1$ can succeed if $m \geq O(\|x^*\|_0 + \|f^*\|_0)\ln((n+m)/(\|x^*\|_0 + \|f^*\|_0))$, which follows by proving that matrix $[A, I]$ satisfies the restricted isometry property when the sensing matrix A is Gaussian matrix with i.i.d entries. [22] improve the results of [28] by allowing $\|f^*\|_0$ to be a positive fraction of the number of observation. Based on the analysis of a extended lasso optimization, [30, 31] further relaxed the sensing matrix A to be Gaussian matrix with i.i.d rows (which is a typical case obeys the extended restricted eigenvalue proposed in [30, 31]), they even allow $\|f^*\|_0$ become arbitrary close to m, however, $m \geq O\|x^*\|_0 \ln(n)\ln(m)$ is necessary for successful recovery. More recently, [23] derives asymptotically similar results as in [28] with specific constant from a more general, convex geometry framework. Recently, [32, 33] study the probabilistic recovery guarantee of a more general $\ell_1$ minimization for varying prior information on $x^*$ and $f^*$:

$$\min_{x,f} \|x\|_1 + \lambda \|f\|_1, \text{ s.t. } Ax + Bf = b \quad (1.4)$$

Where $b = Ax^{(0)} + Bf^{(0)}$, A and B are general matrices, based on the coherence of matrices A and B, the authors in [33] show that recovery of $x^{(0)}$ and $f^{(0)}$ is possible even when the sparsity of $x^{(0)}$ and $f^{(0)}$ scale linearly to the number of measurement m, provided that the signs and supports of $x^{(0)}$ and $f^{(0)}$ satisfy some random assumption. Alternatively, as another line of work on error correction, [13] also proposed to recover $x^*$ from corrupted measurement: $b = Ax^* + f^*$, however, the matrix A in [13] is a tall matrix (m>n), moreover, the recovery method is different from (1.3), in [13], the equation $b = Ax^* + f^*$ is multiplied by a matrix B such that BA=0, and then $f^*$ and $x^*$ is recovered by a $\ell_1$ minimization.

## 1.3 Our contribution

In this paper, we consider the sensing matrix to be sub-gaussian matrix with i.i.d rows, for 2 reasons: firstly because it is a more general case of the Gaussian matrix with i.i.d entries which is frequently studied in literatures, e.g. [22, 28, 29]. Secondly because it is closely related to real applications, for instance, a real world signal y can often be represented as a sparse signal under some orthogonal basis $\Psi$ (e.g., $\Psi$ can be the wavelet basis or Fourier basis, etc.): $y = \Psi x^*$, where $x^*$ here denotes a sparse vector. The measurement vector b is a collection of network data: $b = \Phi y = \Phi \Psi x^*$, then the sensing matrix reads $A = \Phi \Psi$, when $\Phi$ is designed as Gaussian random matrix with i.i.d entries, then A can be naturally interpreted as a subgaussian random matrix with i.i.d rows.

To the problem of compressed sensing with corruption, we are primarily interesting in 2 case: 1) the number of non-zero entries of the corrupted noise vector $\|f^*\|_0$ occupies a positive fraction of the total number of measurement m; 2) $\|f^*\|_0$ becomes arbitrary close to m. These cases can frequently occur in real applications, e.g., [1, 27, 30, 31] .

In the first case, we show that the recovery of $x^*$ and $f^*$ is possible by (1.2) provided that $\|x^*\|_0 \leq O(n/\ln(en/\|x^*\|_0))$, which is also the asymptotically optimal bound, our analysis is based on the generalized restricted isometry property stated in [22] and applying some well-known results on subgaussian random matrix in CS literatures. While in the second case we show that the recovery is still possible as long as $m \geq O\|x^*\|_0 \ln(\|x^*\|_0)$, which is asymptotically better than the bound $m \geq O\|x^*\|_0 \ln(n) \ln(m)$ achieved by recent literatures [30, 31], our analysis is inspired by the elegant golfing scheme proposed in [34]. It is worthy to note that our analysis results still apply when adding bounded, dense noise to the measurement vector b (e.g., the Gaussian measurement noise), or the corrupted noise $f^*$ is transformed under a orthogonal basis.

*Organization of paper.* The organization of the remaining paper is stated as follows, section 2 stated the main results of this paper—theorem 2.1 and theorem 2.2, the proof of theorem 2.1 is given in appendix B, section 3 sketch the proof road map of theorem 2.2, which is a golfing scheme proposed in this paper, with the supporting lemmas of theorem 2.1 stated in appendix C, appendix A provided necessary background on sub-gaussian variable and sub-gaussian matrix on this paper. Section 4 provides numerical experiments which validate the results of theorem 2.1 and theorem 2.2. Finally, section 5 summarizes our finding and future works.

## 2. Main results

This section introduces 2 different setting of $\lambda$ in (1.2) based on some mild prior information of $x^*$ and $f^*$, e.g., a rough estimation of the upper-bound of $\|x^*\|_0$ and $\|f^*\|_0$ and then show the corresponding recovery results, the proofs of these results are referred to section 3 for theorem 2.1 and section 4 for theorem 2.2, respectively.

*Notations.* In this section the sensing matrix A is an $m \times n$ random matrix with independent, isotropic[3], and subgaussian rows with the same subgaussian parameter c, $\hat{A} = [\tilde{A}, I_m]$, where $\tilde{A} = \frac{1}{\sqrt{m}} A$. let $s_x$ denotes an indices set which contains the support set of $x^*$, and $s_f$ denotes an indices set which contains the support set of $f^*$, and $|s_x|$, $|s_f|$ denotes the cardinality of $s_x$, $s_f$, respectively, here $|\cdot|$ denotes the cardinality of $\cdot$ if $\cdot$ denotes a set.

## 2.1 recovery with constant fraction of corruption

Here, we consider a more general, called the stable recovery of (1.2), as described below:
$\min_{x,f} \|x\|_1 + \lambda \|f\|_1$ s.t. $\|Ax + f - b\|_2 \leq \epsilon$ (2.1.1)

In (2.1.1), we assume except the corrupted noise f, the measurement vector b also contains some dense, but bounded noise denoted by v: $b = Ax + f + v$, here in (2.1.1) we assume that $\|v\|_2 \leq \epsilon$. The theoretical guarantee for the recovery performance of (2.1.1) is summarized as in below theorem 2.1.

**Theorem 2.1** *Suppose that A is an $m \times n$ random matrix with independent, isotropic, and subgaussian rows, the signal to be recovered is $x^* \in R^n$, and the measurement vector $b = Ax^* + f^* + v$, where $f^*, w \in R^m$ with $\|v\|_2 \leq \epsilon$. Then by choosing $\lambda = \frac{1}{\sqrt{\ln(en/|s_x|)}}$, the solution $\hat{x}, \hat{f}$ to (2.1.1) obeys*

$\|\hat{x} - x^*\|_2 + \|\hat{f} - f^*\|_2 \leq C_1 \epsilon$ (2.1.2)

*With probability at least $1 - C_2 \exp(-C_3 m)$, for all vector $x^*$ and $f^*$ obeying $\|x^*\|_0 \leq \alpha m / \ln(en/|s_x|)$ and $\|f^*\|_0 \leq \alpha m$. Here, $C_1 \sim C_3$ are some positive constants.*

Theorem 2.1 indicates that stable recovery of (2.1.1) is possible even if $|s_x|$ is as large as $O\left(m/\ln\left(\frac{en}{|s_x|}\right)\right)$ which is also the asymptotically optimal bound. The recovery guarantee (2.1.2) is uniform, which holds for all vector $x^*$ and $f^*$ obeying some sparsity constraints $\|x^*\|_0 \leq \alpha m / \ln(en/|s_x|)$ and $\|f^*\|_0 \leq \alpha m$. The proof of theorem 2.1 is based on the generalized restricted isometry property in [22] and applying some well-known properties of sub-gaussian matrix in CS literatures, see appendix B for details.

### contribution and relevant previous works

Wright et.al [27] show that when A is a i.i.d Gaussian designed matrix as described in their "cross and xx" model, then the exact recovery of (1.3) is possible even when grows arbitrarily close to m, provided that $\|x^{(0)}\|_0$ is sub-linear smaller than m, this upper-bound on $\|x^{(0)}\|_0$ is significantly larger than the bound we obtained in theorem 2.1. [28] and [29] independently show that recovery of (1.3) with $\lambda = 1$ can succeed if $m \geq O(\|x^*\|_0 + \|f^*\|_0)\ln((n+m)/(\|x^*\|_0 + \|f^*\|_0))$, and the sensing matrix A is Gaussian matrix with i.i.d entries, this disallow the corrupted noise occupies a constant fraction of the measurement vector. [22] [1] allows $\|f^*\|_0$ grow linearly to m when A is Gaussian designed or partial BOS, however $m \geq O\|x^*\|_0 \ln(n) \ln(m)$ is necessary for successful recovery, and this lower-bound of m is of course significantly larger than those required in theorem 2.1.

---

[3] See appendix A for the definition of isotropic, subgaussian vector.

The most relevant existing work to theorem 2.1 is theorem 1 in [22], by setting $\lambda = \frac{1}{\sqrt{\ln\left(\frac{n}{m}\right)+1}}$, [22] shows that (1.2) can recover $x^*$ and $f^*$ exactly when A is Gaussian random matrix with i.i.d entries and $\|x^*\|_0 \leq O\left(\frac{m}{\ln\left(\frac{n}{m}\right)+1}\right)$, which is inconsistent[4] with the optimal result: $\|x^*\|_0 \leq O\left(\frac{m}{\ln\left(\frac{n}{\|x^*\|_0}\right)+1}\right)$, where $\|x^*\|_0 > 0$. After scrutinizing the proof of theorem 1.1 in section 2 of [22], we found lemma 2.4 of [22] seems to inaccurately cite the result of theorem 5.2 in [35]: in theorem 5.2 of [35], it states that the Gaussian random matrix A with i.i.d entries satisfies the RIP with a bounded k-RIP constant if $k \leq O\left(\frac{m}{\ln\left(\frac{n}{k}\right)+1}\right)$ which is consistence with the optimal bound, but in lemma 2.4 of [22], it becomes $k \leq O\left(\frac{m}{\ln\left(\frac{n}{m}\right)+1}\right)$. Furthermore, the sensing matrix considered in theorem 2.1 is more general than those in [22].

Finally, it is worthy to mention that the proof of the theorem (see appendix B) also require $\alpha = \frac{|s_f|}{m}$ be a constant sufficiently small to meet some necessary conditions, e.g., the RIP constant of matrix $\left[\frac{1}{\sqrt{m}} A, I_m\right]$ be small enough, which eventually prohibits $\alpha$ to grow arbitrary close to 1. This motivates our works in the next section, where we show that the recovery of $x^*$ and $f^*$ is possible even when $\|f^*\|_0$ grows arbitrarily close to m, provided that $\|x^*\|_0 \leq O\frac{m}{\ln(n)}$, this upper-bound of $\|x^*\|_0$ is only slightly larger than the optimal asymptotical bound.

---

[4] Generally speaking $O\left(\frac{m}{\ln\left(\frac{n}{m}\right)+1}\right)$ is asymptotically smaller than $O\left(\frac{m}{\ln\left(\frac{n}{\|x^*\|_0}\right)+1}\right)$ when m is asymptotically smaller than $O(n)$. Since if the inverse true, say, $O\left(\frac{m}{\ln\left(\frac{n}{m}\right)+1}\right)$ is asymptotically the same as $O\left(\frac{m}{\ln\left(\frac{n}{\|x^*\|_0}\right)+1}\right)$, which can only happen when $\|x^*\|_0 = c_s m$, where $c_s$ is some constant, let $m = c_m \alpha(n)$, where $c_m$ is constant, $\alpha(n)$ is some asymptotical quantity depending on n, then $\|x^*\|_0 \leq O\left(\frac{m}{\ln\left(\frac{n}{\|x^*\|_0}\right)+1}\right)$ implies that $1 \leq c/\left(\ln\left(\frac{n}{c_m \alpha(n)}\right)+1\right)$ where c is some positive constant, this leads to a contradiction when $\lim_{n \to \infty} n/\alpha(n) = 0$, since in this case, the right hand side of the last inequality tends to 0.

## 2.2 recovery with grossly corruption

For convenience, we reformulated (2.1.1) equivalently as:

$$\min\|x\|_1 + \|f\|_1, s.t. A_{|\theta} \begin{bmatrix} x \\ f \end{bmatrix} = b \qquad (2.2.1)$$

Where $A_{|\theta} = [\theta_A A, \theta_I I_m]$, $\theta_A$, $\theta_I$ are some positive constants in (2.2.1), then we have the below theorem shows the recovery performance of (2.2.1):

**Theorem 2.2** let A be an $m \times n$ random matrix with independent, isotropic, and subgaussian rows with the same subgaussian parameter c in (A.2), set $\theta_A = \frac{1}{\sqrt{|s_f|}}$, $\theta_I = c_I \sqrt{\ln(2n/\varepsilon)}$, where $0 < \varepsilon < 1$ is a constant, if $m - |s_f| \geq \max\left\{ c_1 |s_x| \ln(2n/\varepsilon), 3.4 c_I^2 \frac{|s_f|}{m-|s_f|} \ln\left(\frac{2n}{\varepsilon}\right), \frac{8}{3\tilde{c}}(7|s_x| + 2\ln(2\varepsilon^{-1})) \right\}$, with positive constants $c_I$ and $c_1$ satisfies,

$$c_I \geq \max\left\{ \sqrt{8c}, \sqrt{\frac{40c}{|s_f|}} \right\} \qquad (2.2.2a)$$

$$c_1 \geq \max\left\{ 3672c, \frac{94}{\tilde{c}}, \frac{8c_I^2|s_f|}{\tilde{c}(m-|s_f|)}, \frac{158 c_I^2 |s_f|}{m-|s_f|}, \frac{79c}{|s_x|} \right\} \qquad (2.2.2b)$$

Where constant $\tilde{c}$ depends only on the subgaussian parameter c, then with probability $\geq 1 - \sum_{i=1}^{4}(1 - P_{\Delta u}^{(i)}) - \sum_{i=2}^{4}(1 - P_{\Delta h}^{(4)})$, the solution of (2.2.1) can recover $x^*$ and $f^*$ exactly,

Where $P_{\Delta u}^{(i)}, 1 \leq i \leq 4$, $P_{\Delta h}^{(i)}, 2 \leq i \leq 4$ are defined in lemma C.2.1~C.2.4, lemma C.3.1~lemma C.3.3, respectively.

$P_{\Delta u}^{(1)} = 1 - \varepsilon = \qquad , \qquad P_{\Delta u}^{(2)} = (1 - \varepsilon)\left(1 - 2\exp\left(-\frac{\tilde{c}(m-|s_f|)}{8}\right) - \varepsilon\right)$,

$P_{\Delta u}^{(3)} = (1 - \varepsilon)\left(1 - 2\exp\left(-\frac{\tilde{c}(m-|s_f|)}{8}\right) - \varepsilon - \frac{\varepsilon}{n}\right)$, $P_{\Delta u}^{(4)} = (1 - \varepsilon)\left(1 - 2\exp\left(-\frac{\tilde{c}(m-|s_f|)}{4}\right) - 2\exp(-|s_x|) - \varepsilon - \frac{\varepsilon}{n}\right)$.

$P_{\Delta h}^{(2)} = 1 - 2\exp\left(-\frac{\tilde{c}(m-|s_f|)}{8}\right) - \varepsilon$, $P_{\Delta h}^{(3)} = 1 - 2\exp\left(-\frac{\tilde{c}(m-|s_f|)}{8}\right) - \frac{(n+1)\varepsilon}{n}$, $P_{\Delta h}^{(4)} = 1 - 2\exp\left(-\frac{\tilde{c}(m-|s_f|)}{4}\right) - 2\exp(-|s_x|) - \frac{(n+1)\varepsilon}{n}$.

Theorem 2.2 shows that when $|s_x| \leq O(m/\ln(n))$, then (2.2.1) can recover the signal and the corrupted noise exactly when $\frac{|s_f|}{m} < 1$, unlike theorem 2.1, theorem 2.2 allows $\frac{|s_f|}{m}$ becomes arbitrary close to 1, provided that $m - |s_f|$ is asymptotically larger than $|s_x|$ by a $\ln(n)$ factor. We further see that $P_{\Delta u}^{(i)} \to 0, 1 \leq i \leq 4$ and $P_{\Delta h}^{(i)} \to 0, 2 \leq i \leq 4$ provided that the constant $\varepsilon$ is sufficiently small, $|S_x|$, $|S_f|$ and $m - |S_f|$ are sufficiently large.

Finally, it is worthy to mentioned that when b contains some bounded measurement noise which is denoted as v, then the recovery of $x^*$ and $f^*$ can be obtained through the solution of below convex optimization:

$$\min\|x\|_1 + \|f\|_1, s.t. \left\|A_{|\theta}\begin{bmatrix}x\\f\end{bmatrix} - b\right\|_2 \leq \eta$$

Where $\eta$ denotes the upper-bound of the norm of v: $\|v\|_2 \leq \eta$. The stable recovery guarantee for the above convex optimization can also be established by combining the proof of theorem 2.2 and theorem 4.33 in [11].

## Contribution and connections to existing works

Wright et al [27], Li [22] and Ngyuen [1] have shown theoretical recovery guarantee for the weighted $\ell_1$ minimization (2.1.1) for varying sensing matrix A, e.g. the Gaussian designed matrix [22, 27] or Partial BOS chap. 12 [11]. However, their results required significantly more number of measurements than those required in our theorem 2.2, for example, Li [22] and Ngyuen [1] require that $m \geq O(\|x^*\|_0 \ln(m)\ln(n))$ which is asymptotically larger than the upper-bound of m stated in theorem 2.2 by a $\ln(n)$ factor. Furthermore, there are some extra random assumptions on the supports or signs of $x^*$ and $f^*$ in order to achieve the analytic results in [27] [1, 22].

The most closely related works to theorem 2.2 are stated in [30, 31], where the authors propose an extended lasso optimization to recover $x^*$ and $f^*$ from a noisy measurement vector $b = Ax^* + \sqrt{m}f^* + w$ as stated in (2.2.3), where A is a standard Gaussian designed matrix, entries in the m dimensional vector w are i.i.d Gaussian random variables with variance $\sigma^2$.

$$\min_{x,f} \frac{1}{2m}\|b - Ax - \sqrt{m}f\|_2^2 + \lambda_x\|x\|_1 + \lambda_f\|f\|_1 \qquad (2.2.3)$$

Where the parameters $\lambda_x$, $\lambda_f$ are depended on $\sigma$, m, n as suggested by theorem 2 of [30], then the recovery error of $x^*$ and $f^*$ is proportional to $\sigma$, as long as $m \geq O(\|x^*\|_0 \ln(m)\ln(n))$, theorem 3 of [30] further shows that this lower-bound is indeed optimal for stable recovery of (2.2.3), one fundamental drawback of the extended lasso (2.2.3) comparing to the proposed (2.2.1) is that the lower bound for the number of measurement m in (2.2.3) is asymptotically larger than those of (2.2.1) by a $\ln(n)$ factor as suggested by theorem 2.2.

Moreover, it is difficult to evaluate the performance of the extended lasso (2.2.3) when $\sigma \to 0$ (e.g. when w is close to be a 0 vector), although theorem 2 in [30] implies that when $\sigma = 0$, (2.2.3) recover $x^*$ and $f^*$ exactly by its solution, we found that when $\sigma = 0$, one has $\lambda_x = \lambda_f = 0$, then (2.2.3) is degenerated into a least-square problem:

$$\min_{x,f}\|b - Ax - \sqrt{m}f\|_2^2 \qquad (2.2.4)$$

Since the solution of x, f to the linear equations system $b - Ax - \sqrt{m}f = 0$ is not unique, it is difficult to see how the extended lasso optimization (2.2.3) can achieve exact recovery in this case.

# 3. Proof of theorem 2.2

This section provides a brief roadmap for the proof of theorem 2.2, with the supporting lemmas provided in appendix B, and appendix A provides known results on sub-gaussian random variable which are necessary in this paper.

*Notations* [m] denotes the indices set $\{1, \ldots, m\}$ if m represents a positive integer, and $i + [m] = \{i + 1, \ldots, i + m\}$ denotes the indices set [m] shifted by i, where i denotes an integer, given an indices set S such that $S \subseteq [m]$, we denote the complement set of S by $S^c$ or $[m]\backslash S$ in this paper. Let $\sigma_x = sgn(x^*(s_x)), \sigma_f = sgn(f^*(s_f))$ denotes the corresponding sign vector of $x^*, f^*$ respectively.

Firstly, we state a sufficient and necessary condition for the exact recovery of $\ell_1$ minimization, which is called the dual certification condition [36].

**Lemma 3.1** ( *dual certification, theorem 4 in [36]*) Given a matrix $A \in R^{m \times N}$, a vector $x \in R^N$ with support S is the unique minimizer of $\|z\|_1$ subject to $Az = Ax$ if the following condition holds:
There exist a vector $h \in R^m$, such that,
$$A([m], s)^T h = sgn(x(s)), \|A([m], s^c)^T h\|_\infty < 1 \quad (3.1)$$
And matrix $A([m], s)$ is full rank.

A straightforward application of lemma 3.1 to our algorithm (2.2.1) gives the below lemma.
**Lemma 3.2** $x^*$ and $f^*$ is the unique solution of (2.2.1) if and only if the following condition holds:
There exists a vector $h \in C^m$ such that,
$$\begin{cases} \theta_A A^T(s_x, [m])h = \sigma_x, \theta_I h(s_f) = \sigma_f \\ \|\theta_A A^T(s_x^c, [m])h\|_\infty < 1, \|\theta_I h(s_f^c)\|_\infty < 1 \end{cases} \quad (3.2)$$

And matrix $B = [\theta_A A([m], s_x), \theta_I I_m([m], s_f)]$ is full rank.
Proof: this follows from a direct application of lemma 3.1. ∎

According to lemma 3.2, to prove (2.2.1) can correctly recover $x^*, f^*$, our goal is therefore to construct a viable vector $h \in R^m$, such that (3.2) in lemma 3.2 holds. To this end we'll construct such h via a simple golfing scheme as stated below:

*Golfing scheme to construct h:*
0. "initialization": let $h = 0$.
1. "Hitting $\sigma_f$": Construct a $\Delta h^{(1)}$, such that $\Delta h^{(1)}(s_f) = \frac{1}{\theta_I}\sigma_f, \Delta h^{(1)}(s_f^c) = 0$, set $h = h +$

$\Delta h^{(1)}$, $\Delta u^{(1)} = A_{|\theta}^T \Delta h^{(1)}$, $u^{(1)} = A_{|\theta}^T h$ and $w^{(1)} = \sigma_x - u^{(1)}(s_x)$.

2. "Approaching $\sigma_x$": Choose a subset $\Lambda_1 \subset [m]\backslash s_f$ (with $|\Lambda_1| = \frac{m-|s_f|}{2}$) as the indices set corresponding to the smallest (measured in absolute value) of $A_{|\theta}([m]\backslash s_f, s_x) w^{(1)}$. Construct a vector $\Delta h^{(2)}$, such that $\Delta h^{(2)}(\Lambda_1) = \frac{2|s_f|}{m-|s_f|} A_{|\theta}(\Lambda_1, s_x) w^{(1)}, \Delta h^{(2)}(\Lambda_1^c) = 0$, let $h = h + \Delta h^{(2)}$, $\Delta u^{(2)} = A_{|\theta}^T \Delta h^{(2)}$, $u^{(2)} = A_{|\theta}^T h$ and $w^{(2)} = \sigma_x - u^{(2)}(s_x)$.

3. "Approaching $\sigma_x$": Similar to the step 2, choose a subset $\Lambda_2 \subset [m]\backslash s_f$ (with $|\Lambda_2| = \frac{m-|s_f|}{2}$) as the smallest (measured in absolute value) indices set of $A_{|\theta}([m]\backslash s_f, s_x) w^{(2)}$. Construct a vector $\Delta h^{(3)}$, such that $\Delta h^{(3)}(\Lambda_2) = \frac{2|s_f|}{m-|s_f|} A_{|\theta}(\Lambda_2, s_x) w^{(2)}, \Delta h^{(3)}(\Lambda_2^c) = 0$, let $h = h + \Delta h^{(3)}$, $\Delta u^{(3)} = A_{|\theta}^T \Delta h^{(3)}$, $u^{(3)} = A_{|\theta}^T h$ and $w^{(3)} = \sigma_x - u^{(\ \ )}(s_x)$.

4. "Hitting $\sigma_x$": Let $\Lambda_3 = [m]\backslash s_f$, construct a vector $\Delta h^{(4)}$, such that $\Delta h^{(4)}(\Lambda_3) = A_{|\theta}^\dagger(\Lambda_3, s_x) w^{(3)}$, $\Delta h^{(4)}(\Lambda_3^c) = 0$, let $h = h + \Delta h^{(4)}$, $\Delta u^{(4)} = A_{|\theta}^T \Delta h^{(4)}$ and $u = A_{|\theta}^T h$ where $A_{|\theta}^\dagger(\Lambda_3, s_x) = A_{|\theta}(\Lambda_3, s_x) \left( A_{|\theta}^T(s_x, \Lambda_3) A_{|\theta}(\Lambda_3, s_x) \right)^{-1}$ denotes the Penron-Moore inverse of matrix $A_{|\theta}(\Lambda_3, s_x)$.

We'd like to add some comments on our golfing scheme before we show its validity, which are stated below:

- Constructing $\Delta h^{(2)}$ and $\Delta h^{(3)}$ using a particular support sets $\Lambda_1$ and $\Lambda_2$ as described in step 2 and step 3 is crucial to achieve the bound: $m \geq |s_x| \ln(n)$, otherwise, the upper-bound for m might increase by a $\ln(n)$ factor as in [22].

- The last golfing step—step 4 plays a similar role as the so-called "inexact duality conditions" frequently used in [11, 17, 22, 34] for the convenience of showing the validity of the golfing schemes and establishing the stable recovery of the $\ell_1$ minimization, the "inexact duality conditions" are generally derived by the primal problem using somewhat tricky skills which are less intuitive to the readers, interestingly, we find all of the "inexact duality conditions" in literatures [22, 34, 37] can be equivalently replaced by a additional, straightforward golfing step similar as step 4 in this paper. For simplicity, we should prove the validity of our golfing scheme by verifying the dual certification (3.2) directly.

- In some applications, e.g., when the sensing matrix A is partial BOS, since it might take $> O(1)$ golfing steps (e.g. $O(\ln(n))$ steps as in [11, 22, 37]) to hit $\sigma_x$, to ensure $\|\theta_I h(s_f^c)\|_\infty < 1$ during the golfing scheme, the authors in [11, 22, 37] use disjoint sub-matrices of $A_{|\theta}([m]\backslash s_f, [n]])$ in each golfing steps approaching $\sigma_x$, since in this paper, it requires only 3 steps to hit $\sigma_x$, for the simplicity of the proof, we don't require $\Lambda_1 \sim \Lambda_3$ to be disjoint.

- By slightly modifying of the golfing above, similar the conclusion of theorem 2.2 also hold when the sensing matrix A is partial BOS, with additional condition that the support of $x^*$ is uniformly at random.

To proof the h achieved by the golfing scheme above is indeed a viable vector, we have to show below (3.3) holds with high probability:

$h(s_f) = \sigma_f$     (3.3.a)
$u(s_x) = \sigma_x$     (3.3.b)
$\|u(s_x^c)\|_\infty < 1$     (3.3.c)
$\|h(s_f^c)\|_\infty < 1$     (3.3.d)

Where (3.3.a) follows from step 1 and (3.3.b) follows from step 2 ~ step 4 in the above golfing scheme. The remaining goal is thus to prove (3.3.c~3.3.d) holds with high probability.

Since $h = \sum_{i=1}^{4} \Delta h^{(i)}$ and $u = \sum_{i=2}^{4} \Delta u^{(i)}$, to show (3.3.c~3.3.d), it's sufficient to show:

$$\left\|\Delta u^{(1)}(s_x^c)\right\|_\infty < \frac{1}{\sqrt{2}}, \left\|\Delta u^{(i)}(s_x^c)\right\|_\infty < \frac{1}{3}(1 - 1/\sqrt{2}), 1 \le i \le 4 \quad (3.4.a)$$

$$\left\|\Delta u^{(i)}(n + s_f^c)\right\|_\infty = \theta_I\left\|\Delta h^{(i)}(s_f^c)\right\|_\infty < \frac{1}{3}, 2 \le i \le 4 \quad (3.4.b)$$

In this paper, we choose $\theta_A = \frac{1}{\sqrt{|s_f|}}$, $\theta_I = c_I\sqrt{\ln(2n/\varepsilon)}$, assuming that A be an $m \times n$ random matrix with independent, isotropic, and subgaussian rows with the same subgaussian parameter c as described in appendix A.

The remaining part of the proof is organized as following, in order to achieved (3.4.a~3.4.b) with high probability, firstly we bound $\left\|w^{(i)}\right\|_2, 1 \le i \le 3$ in section C.1, and then we bound $\left\|\Delta u^{(i)}(s_x^c)\right\|_\infty, \theta_I\left\|\Delta h^{(i)}(s_f^c)\right\|_\infty$ stated in (3.4.a~3.4.b) in section C.2 and section C.3, respectively.

Finally, we prove that B in lemma 3.2 is full rank in appendix C.4, by putting together the conclusions in C.1~C.4, it eventually leads to a natural proof for theorem 2.2 in section C.5.

## 4. Experiments

In this section, we provide simulation experiments to illustrate the theoretical results suggested by theorem 2.1 and theorem 2.2 simulations are performed for a range of parameters $(n, m, |s_x|, |s_f|)$.

Where $n = \{128, 256, 512\}$, $\vartheta_m = \frac{m}{n} = \{0.1, 0.2, \ldots, 1\}$ denotes the ratio of observation, $\vartheta_f = \frac{|s_f|}{m} = \{0.1, 0.2, \ldots, 0.5\}$ denotes the fraction of corrupted noise, $|s_x| = \lfloor 0.2n/\ln(0.2n) \rfloor + 1$ which is asymptotically similar to $|s_x|$ as suggested in theorem 2.1 and theorem 2.2.

The sensing matrix A is chosen as Gaussian matrix with independent, isotropic rows as suggested previously, $x^*$ and $f^*$ are generated with random support sets and the magnitude of non-zero elements in $x^*$ and $f^*$ are obey normal distribution where the variance of non-zeros elements in $f^*$ is 100 times larger than those in $x^*$.

The recovery error is measured by relative error:

$$\text{RE} = \frac{\left\| \begin{bmatrix} x^* \\ f^* \end{bmatrix} - \begin{bmatrix} \hat{x} \\ \hat{f} \end{bmatrix} \right\|_2}{\left\| \begin{bmatrix} x^* \\ f^* \end{bmatrix} \right\|_2} \times 100\% \qquad (4.1)$$

## 4.1 illustrating theorem 2.1

In this section, $\begin{bmatrix} \hat{x} \\ \hat{f} \end{bmatrix}$ is obtained by the solution of (2.1.1) with parameter $\lambda = \frac{1}{\sqrt{\ln(n/|s_x|)}}$ as suggested in theorem 2.1. In practice, since $s_x$ is a non-empty indices set which contains the support of $x^*$, thus $|s_x|$ is only a rough estimation of the upper bound of $\|x^*\|_0$, which is usually not difficult to obtain and therefore the only parameter $\lambda$ in (2.1.1) is not hard to define.

The results are summarized in heat-maps of figure (4.1.1). We assert accurate recovery is succeed if the RE value in (4.1) is smaller than $10^{-8}$, each element in the each heat-map indicates the average recovery succeed rate of a particular setting of parameters $(n, m, |s_x|, |s_f|)$, where the average succeed rate is calculated by 100 independent runs.

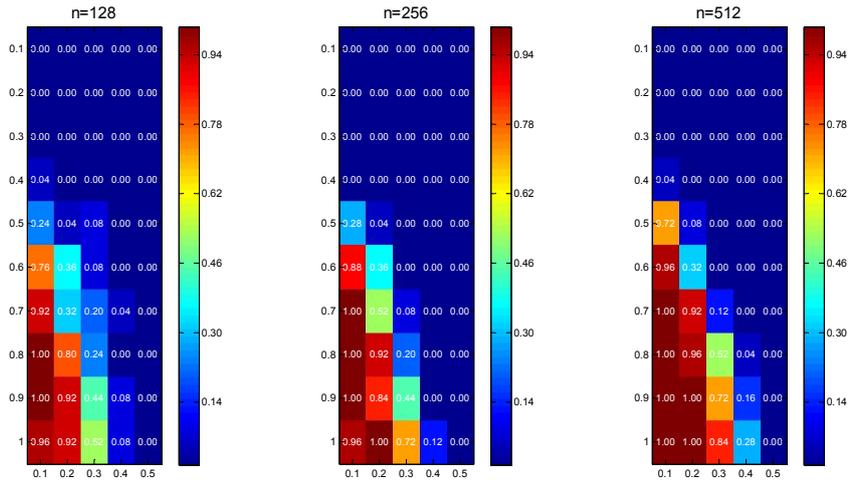

Figure (4.1.1) the recovery performance of (2.1.1) when $\lambda = \frac{1}{\sqrt{\ln(n/|s_x|)}}$. Element in each heat-map indicates the average recovery succeed rate of 100 independent runs based on the simulation data (see the description of the simulation data in above text), with a particular parameters setting of $(n, \vartheta_m, \vartheta_f)$, the vertical axis of the heat-maps indicate different values for $\vartheta_m$ (the measurement rate), the horizontal axis of the heat-maps indicate different values for $\vartheta_f$ (the corruption rate), and the titles of the heat-maps indicate n (the dimension of input signal to be recovered). E.g., the first row, first column element of the left-most heat-map indicate the average recovery successful rate of 100 independent runs based on the simulation data, when $n = 128$, $\vartheta_m = 0.1$ and $\vartheta_f = 0.1$.

As we can see from figure (4.1.1), when the measurement rate $\vartheta_m \geq 0.7$ and the corruption rate $\vartheta_f \leq 0.1$, the recovery succeed rate of (2.1.1) is close to one in all case (n=128, 256, 512) which verifies the conclusion of theorem 2.1, as a generally trend, the succeed rate gradually decrease when the $\vartheta_m$ decrease and $\vartheta_f$ increase.

## 4.2 illustrating theorem 2.2

In this section, $\begin{bmatrix}\hat{x}\\\hat{f}\end{bmatrix}$ is obtained by the solution of (2.2.1) with parameters $\theta_A = \frac{1}{\sqrt{|s_f|}}$, $\theta_I = c_I\sqrt{\ln(2n/\varepsilon)}$ as suggested in theorem 2.2, we set $|s_f| = 0.1m$ (which means we should allow no more than 10% corruption), $c_I = 2$ as suggested in (2.2.1) because the parameter c of Gaussian random variable is 1/2. Finally, we set $\varepsilon = 0.01$, we find the performance of (2.2.1) is insensitive to $\varepsilon$ when we choose $0 < \varepsilon \leq 0.01$, because in this case $\sqrt{\ln(n/\varepsilon)}$ increase very slowly as $\varepsilon$ decreases. In practice, since $s_f$ represents a non-empty indices set that contains the support of $f^*$, $|s_f| > 0$ defining $\theta_A$ is only a rough estimation of the upper-bound of $\|f^*\|_0$, which is not hard to obtained in real application.

The results are summarized in heat-maps of figure (4.2.1) , the meaning of figure (4.2.1) is defined similarly as in section 4.1.

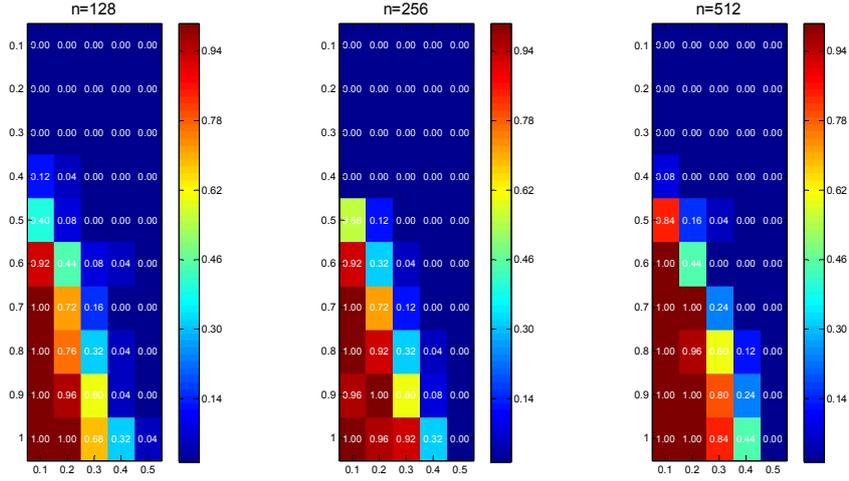

Figure (4.2.1) the recovery performance of (2.2.1) when $\theta_A = \frac{1}{\sqrt{|s_f|}}$, $\theta_I = c_I\sqrt{\ln(2n/\epsilon)}$, Element in each heat-map indicates the average recovery succeed rate of 100 independent runs based on the simulation data (see the description of the simulation data in above text), with a particular parameters setting of $(n, \vartheta_m, \vartheta_f)$, the vertical axis of the heat-maps indicate different values for $\vartheta_m$ (the measurement rate), the horizontal axis of the heat-maps indicate different values for $\vartheta_f$ (the corruption rate), and the titles of the heat-maps indicate n (the dimension of input signal to be recovered). E.g., the first row, first column element of the left-most heat-map indicate the average recovery successful rate of 100 independent runs based on the simulation data, when $n = 128$, $\vartheta_m = 0.1$ and $\vartheta_f = 0.1$.

Figure (4.2.1) shows that the recovery succeed rate are close to be 1 when the measurement rate measurement rate $\vartheta_m \geq 0.7$ and the corruption rate $\vartheta_f \leq 0.1$ when n=128, 256 and 512, which verifies the conclusion of theorem 2.2, similar to the results in figure (4.1.1), the succeed

recovery rate decreases gradually as the decrement of measurement rate and the increment of the corruption rate. Although it allows $\vartheta_f$ become arbitrary close to 1 in theorem 2.2, notice that as $|s_f|$ approaches to m it also requires $|s_x|$ decrease to 0 more rapidly than $m - |s_f|$. Since in our experiment in this section, we set $|s_x| = \lfloor 0.2n/\ln(0.2n) \rfloor + 1$, which is always bounded away from 0, this is why we observe that the recovery is always failed in figure (4.2.1) when $\frac{|s_f|}{m} > 0.5$. To ensure the successful recovery of (2.3.1) with higher corruption ratio, the upper-bound on the sparsity of $x^*$ should become tighter accordingly.

# 5. Conclusions and future work

In summary, this paper proves that the signal $x^*$ can be recover from a corrupted linear measurement $b = Ax^* + f^*$ within a reweighted $\ell_1$ minimization framework (1.2), when a constant fraction of the measurement b is corrupted if $\|x^*\|_0 > 0$ is less than an optimal upper bound $O(n/\ln(en/\|x^*\|_0))$, when the upper-bound of $\|x^*\|_0$ is slightly larger than the optimal upper-bound, say, $\|x^*\|_0 \geq O(n/\ln(n))$, then the recovery is possible even when the corruption ratio grows arbitrarily close to 1.

Recently [17]show that one can recover a low-rank matrix $L_0 \in R^{n_1 \times n_2}$ from its grossly corrupted observation $M = L_0 + S_0$ by solving the following nuclear-norm optimization:
$$\min_{L,S} \|L\|_* + \lambda \|S\|_1, s.t. L + S = M \qquad (5.1)$$

The author in [17] prove (via also a golfing scheme inspired from [34]) that the solution of (5.1) can recover $L_0$ even when almost all elements in $L_0$ are arbitrarily corrupted by $S_0$, provided that the rank of $L_0$: $\text{rank}(L_0) \leq O\left(\frac{n}{\ln^2(n)}\right)$, where $n = \max\{n_1, n_2\}$, however, we notice that in the numerical results provided by [17], the recovery results are quite well when $\text{rank}(L_0) \leq O(n)$ and a constant fraction of elements in $L_0$ are corrupted, motivated by the results of theorem 2.2 in this paper, this drives us to believe that the upper-bound on $\text{rank}(L_0)$ might be improved to be $O\left(\frac{n}{\ln(n)}\right)$ through a modified golfing scheme, such work are ongoing.

# Acknowledgement

The author wish to thank his wife Haixia Yu for her mental as well as material supports over the past 8 years.

# Appendix A—subgaussian variables

In this paper, we are primarily interested in the case where the measurement matrix A in is row independent subgaussian matrix. To gain a good understanding of matrix A, we should firstly introduce the definitions such as subgaussian variable and subgaussian vector which constructed

A, and then we summarize some useful properties of A which used in the proofs of this paper.

**Definition A.1** (subgaussian variable [11]) A random variable X is called subgaussian if there exist constants $\beta, \kappa > 0$ such that
$$Prob(|X| \geq t) \leq \beta e^{-\kappa t^2} \quad \text{for all } t > 0 \quad \quad (A.1)$$
A mean-0 subgaussian variable might be defined equivalently via its Laplacian transformed as stated in the following proposition A.

**Proposition A.1** (proposition 7.24 of [11]). Let X be a random variable.
(a) If X is subgaussian with $E[X] = 0$, then there exists a constant $c$ (depending only on $\beta, \kappa$) such that $E[exp(\theta X)] \leq exp(c\theta^2)$ for all $\theta \in R$. (A.2.a)
(b) Conversely, if (A.2.a) holds, then $E[X] = 0$ and X is subgaussian with parameters $\beta = 2$ and $\kappa = \frac{1}{4c}$.

We now extend the definition of subgaussian variable to the definition of subgaussian random vector, as stated below,

**Definition A.2** (definition 9.4 in [11], subgaussian vector) Let Y be a random vector on $R^n$.
(a) If $E[|\langle Y, x \rangle|^2] = \|x\|_2^2$ for all $x \in R^N$, then Y is called isotropic.
(b) If, for all $x \in R^n$ with $\|x\|_2 = 1$, the random variable $\langle Y, x \rangle$ is subgaussian with subgaussian parameter c being independent of x, that is,
$E[exp(\theta \langle Y, x \rangle)] \leq exp(c\theta^2)$ for all $\theta \in R, \|x\|_2 = 1$ (A.2.b)
Then Y is called a subgaussian random vector.

It's easy to see that, if Y is a subgaussian random vector with parameter c, then each element of Y can be treated as a subgaussian random variable with parameter c.
The below lemmas state useful properties of subgaussian variable which are necessary in our proofs of this paper.

**Lemma A.1** (theorem 7.27 of [11]) let $X_1, \cdots, X_M$ be a sequence of independent mean-zero subgaussian random variables with subgaussian parameter c in (A.2.a). For $a \in R^M$, the random variable $Z \coloneqq \sum_{i=1}^{M} a_i X_i$ is subgaussian, i.e.,
$E[exp(\theta Z)] \leq exp(c\|a\|_2^2 \theta^2)$ (A.3)
And,
$prob(|Z| \geq t) \leq 2exp\left(-\frac{t^2}{4c\|a\|_2^2}\right)$ for all $t > 0$ (A.4)

**Lemma A.2** (lemma 9.8 of [11]) let A be an $m \times n$ random matrix with independent, isotropic, and subgaussian rows with the same subgaussian parameter c in (A.2.b). Then, for all $x \in R^n$ and every $t \in (0,1)$,
$prob(|m^{-1}\|Ax\|_2^2 - \|x\|_2^2| \geq t\|x\|_2^2) \leq 2exp(-\tilde{c}t^2 m)$ (A.5)
where $\tilde{c}$ depends only on c.

**Lemma A.3** *(theorem 9.9 of [11]) Suppose that an $m \times n$ random matrix A is drawn according to a probability distribution for which the concentration inequality (A.6) holds, that is, for $t \in (0,1)$,*
$prob(|\|Ax\|_2^2 - \|x\|_2^2| \geq t\|x\|_2^2) \leq 2exp(-\tilde{c}t^2m)$ *for all* $x \in R^n$   (A.6)
*for $s \subset [n]$ and $\delta, \varepsilon \in (0,1)$, if*

$$m \geq \frac{2}{3\tilde{c}}\delta^{-2}(7|s| + 2ln(2\varepsilon^{-1})) \qquad (A.7)$$

*then with probability at least $1 - \varepsilon$, $\left\|A^T(s,[m])A([m],s) - I_{|s|}\right\|_{2\to 2} < \delta$.*

## Appendix B—Proof of theorem 2.1

**Definition B.1** *(generalized restricted isometry property (RIP) [22]) For any matrix $\Phi \in R^{L\times(n+m)}$, define the RIP-constant $\delta_{s_1,s_2}$ by the infimum value of $\delta$ such that*

$$(1-\delta)(\|x\|_2^2 + \|f\|_2^2) \leq \left\|\Phi\begin{bmatrix}x\\f\end{bmatrix}\right\|_2^2 \leq (1+\delta)(\|x\|_2^2 + \|f\|_2^2) \qquad (B.1)$$

*Holds for any $x \in R^n$ with $\|x\|_0 \leq s_1$ and $f \in R^m$ with $\|f\|_0 \leq s_2$.*

**Lemma B.1** *(lemma 2.3 of [22]) Suppose $\Phi \in R^{L\times(n+m)}$ with RIP-constant $\delta_{2s_1,2s_2} < \frac{1}{18}$ ($s_1, s_2 > 0$) and $\lambda$ in (2.1.1) is between $0.5\sqrt{\frac{s_1}{s_2}}$ and $2\sqrt{\frac{s_1}{s_2}}$. Then for any $x \in R^n$ with $\|x\|_0 \leq s_1$, any $f \in R^m$ with $\|f\|_0 \leq s_2$, and any $v \in R^m$ with $\|v\|_2 \leq \epsilon$, the solution to the optimization problem (2.1.1) satisfies*

$$\|\hat{x} - x\|_2 + \|\hat{f} - f\|_2 \leq \frac{4\sqrt{13+13\delta_{2s_1,2s_2}}}{1-9\delta_{2s_1,2s_2}}\epsilon \qquad (B.2)$$

**Lemma B.2** *(subgaussian RIP, theorem 5.65 in [15]) let A be $m \times n$ sub-gaussian matrix with independent, isotropic rows, then the normalized matrix $\tilde{A} = \frac{1}{\sqrt{m}}A$ satisfies the following for every sparsity level $1 \leq k \leq n$ and every number $\delta \in (0,1)$: if $m \geq C_1\delta^{-2}kln(en/k)$ then with probability at least $1 - 2exp(-C_2\delta^2m)$, $(1-\delta)\|x\|_2^2 \leq \|\tilde{A}x\|_2^2 \leq (1+\delta)\|x\|_2^2$ holds for any x with $\|x\|_0 \leq k$, where constants $C_1, C_2$ depend only on the sub-gaussian property of A.*

**Lemma B.3** *(theorem 5.39 of [15]) Let A be an $N \times n$ matrix whose rows are independent, sub-gaussian isotropic random vectors in $R^n$. Then for every $t \geq 0$, with probability at least $1 - 2exp(-C_1t^2)$ one has,*
$\sqrt{N} - C_2\sqrt{n} - t \leq \sigma_{min}(A) \leq \sigma_{max}(A) \leq \sqrt{N} + C_2\sqrt{n} + t$   (B.3)
*Where $C_1 \sim C_2$ are positive constants depend only the sub-gaussian property of A, here $\sigma_{min}(\cdot)$ and $\sigma_{max}(\cdot)$ denote the minimum and maximum singular value of matrix.*

**Proof of theorem 2.1** *(which makes necessary modifications upon the proof of theorem 1.1 in [22].)*

Set $|s_x| = \left\lfloor \alpha \frac{m}{\ln(en/|s_x|)} \right\rfloor$ and $|s_f| = \lfloor \alpha m \rfloor$ where $\alpha$ is a small constant be specified later, to prove the theorem, we'd like to bound the RIP-constant $\delta_{2|s_x|,2|s_f|}$ for the $m \times (m+n)$ matrix $\Phi = \left[\frac{1}{\sqrt{m}}A, I_m\right]$ and then apply lemma B.2. Let $T, V$ be any indices sets satisfying:

$\text{supp}(x^*) \subseteq T,\ |T| = 2|s_x|$ and $\text{supp}(f^*) \subseteq V,\ |V| = 2|s_f|$ \hfill (B.4)

then, one has,

$$\left\| \Phi \begin{bmatrix} x^* \\ f^* \end{bmatrix} \right\|_2^2 = \left\| \frac{1}{\sqrt{m}} A x^* \right\|_2^2 + \|f^*\|_2^2 + 2 \langle \frac{1}{\sqrt{m}} A(V, T) x^*, f^* \rangle \quad \text{(B.5)}$$

By lemma B.2, assuming $\alpha \leq C_1^{-1}\delta^2$, then with probability $\geq 1 - 2\exp(-C_2\delta^2 m)$,

$$(1-\delta)\|x\|_2^2 \leq \left\| \frac{1}{\sqrt{m}} A x \right\|_2^2 \leq (1+\delta)\|x\|_2^2 \quad \text{(B.6)}$$

Holds uniformly for any such $T$ and $x^*$ satisfy (B.4), here constants $C_1, C_2$ have the same meaning with those in lemma B.2, which depend only on the sub-gaussian property of matrix A.

Fixing a pair of $T$ and $V$ which satisfy (B.4), and we'd like to bound $\left\|\frac{1}{\sqrt{m}}A(V,T)\right\|_{2\to 2}$. According to lemma B.3,

$$\left\|\frac{1}{\sqrt{m}}A(V,T)\right\|_{2\to 2} \leq \frac{1}{\sqrt{m}}\left(C_3\sqrt{2|s_x|} + \sqrt{2|s_f|} + \sqrt{\delta^2 m}\right) \leq (1+C_3)\sqrt{2\alpha} + \delta \quad \text{(B.7)}$$

Holds with probability at least $1 - 2\exp(-C_4\delta^2 m)$, here $C_3, C_4$ have the same meaning of $C_2, C_1$ in lemma B.3, which depend only on the sub-gaussian property of A. Then applying a union bound, one has with probability at least

$$1 - 2\exp(-C_4\delta^2 m)\binom{n}{2|s_x|}\binom{m}{2|s_f|} \quad \text{(B.8)}$$

(B.7) holds uniformly for any aforementioned $V$ and $T$ satisfying (B.4). Since $2|s_x| \leq \alpha m/\ln(en/|s_x|)$, one has $2|s_x|\ln\left(\frac{em}{2|s_x|}\right) \leq \alpha_1 m$ with $\alpha_1 = 2\alpha$, which implies that $\binom{n}{2|s_x|} \leq \exp(\alpha_1 m)$, similarly, because $2|s_f| \leq \alpha m$, we have $2|s_f|\ln\left(\frac{em}{2|s_f|}\right) \leq \alpha_2 m$, with $\alpha_2 = 2\ln\left(\frac{em}{2|s_f|}\right)\alpha$, where $\alpha_1, \alpha_2$ depends only on $\alpha$ and $\alpha_1 \to 0, \alpha_2 \to 0$ as $\alpha \to 0$, which implies $\binom{m}{2|s_f|} \leq \exp(\alpha_2 m)$. Therefore, (B.7) holds uniformly for any such $V$ and $T$ with probability at least

$$1 - 2\exp\left(-(C_4\delta^2 - \alpha_1 - \alpha_2)m\right) \quad \text{(B.9)}$$

Combining (B.5~B.7), we have,

$(1-\delta)\|x^*\|_2^2 + \|f^*\|_2^2 - \left(\delta + (1+C_3)\sqrt{2\alpha}\right)\|x^*\|_2\|f^*\|_2 \leq \left\|\Phi\begin{bmatrix}x^*\\f^*\end{bmatrix}\right\|_2^2 \leq (1+\delta)\|x^*\|_2^2 +$

$\|f^*\|_2^2 + \left(\delta + (1+C_3)\sqrt{2\alpha}\right)\|x^*\|_2\|f^*\|_2$ \hfill (B.10)

Holds uniformly for any such $x^*, f^*, T$ and $V$ satisfying (B.4) with probability at least $1 - 2\exp(-C_2\delta^2 m) - 2\exp\left(-(C_4\delta^2 - \alpha_1 - \alpha_2)m\right)$. After choosing appropriate $\delta$ and letting $\alpha$ sufficiently small[5], we have $\delta_{2|s_x|,2|s_f|} < \frac{1}{18}$, with probability $\geq 1 - c_1\exp(-c_2 m)$, where

---

[5] E.g., one can choose $\alpha = c_0\delta^2$ where $c_0$ is some constant depends on the subgaussian property of A, firstly,

$c_1 \sim c_2$ are some positive constant.

Finally, under the assumption that $\alpha \frac{m}{\ln(en/|s_x|)} \geq 1$, we have $|s_x| = \left\lfloor \alpha \frac{m}{\ln(en/|s_x|)} \right\rfloor > 0$ and $|s_f| = \lfloor \alpha m \rfloor > 0$ and $\frac{1}{\sqrt{\ln(en/|s_x|)}} \in (0.5, 2)\sqrt{\frac{|s_x|}{|s_f|}}$, then theorem 2.1 is a directly conclusion of lemma B.1. ∎

# Appendix C—Supporting lemmas of theorem 2.2

In this section A is a row i.i.d matrix whose rows are isotropic, subgaussian row vectors, $\theta_A$ and $\theta_I$ are defined as in theorem 2.2, and we assume $m - |s_f| \geq c_1|s_x|\ln(2n/\varepsilon)$ as indicated in theorem 2.2.

## C.1 Bounding $\|w^{(i)}\|_2$

**Lemma C.1.1** (bounding $\|\Delta u^{(1)}([n])\|_\infty$) if $c_I \geq \sqrt{8c}$, then one has,

$$prob\left(\|\Delta u^{(1)}([n])\|_\infty < \frac{1}{\sqrt{2}}\right) \geq 1 - \varepsilon \qquad (C.1.1)$$

Proof: Since $\Delta u^{(1)}(i) = \langle A(s_f, i), \frac{\theta_A}{\theta_I}\sigma_f \rangle, i \in [n]$, where $\langle \cdot, \cdot \rangle$ denotes inner product between 2 vectors, because elements in vector $A(s_f, i)$ are independent subgaussian mean-0 vector with parameter c, according to (A.4) in lemma A.1 and letting $t = \frac{1}{\sqrt{2}}$ yields,

$$prob\left(|\Delta u^{(1)}(i)| \geq \frac{1}{\sqrt{2}}\right) \leq 2\exp\left(-\frac{c_I^2 \ln(2n/\varepsilon)}{8c}\right) \qquad (C.1.2)$$

(C.1.2) implies that if $c_I \geq \sqrt{8c}$, then $|\Delta u^{(1)}(i)| \geq \frac{1}{\sqrt{2}}$ holds with probability at least $\frac{\varepsilon}{n}$, consequently, applying a union bound yields $prob\left(\|\Delta u^{(1)}([n])\|_\infty < \frac{1}{\sqrt{2}}\right) \geq 1 - \varepsilon$ which proves the conclusion of the lemma. ∎

**Lemma C.1.2** (bounding $\|w^{(1)}\|_2$) if $c_I \geq \sqrt{8c}$, then one has,

$$prob\left(\|w^{(1)}\|_2 < (1 + 1/\sqrt{2})\sqrt{|s_x|}\right) \geq 1 - \varepsilon \qquad (C.1.2)$$

Proof: the conclusion of this lemma follows from the triangle inequality and (C.1.1) in lemma C.1.1. ∎

---

letting $\delta$ sufficiently small, then from (B.6) and (B.10) one concludes that $\delta_{2|s_x|,2|s_f|} < \frac{1}{18}$ holds as long as $\delta$ be sufficiently small. Secondly, since $\alpha_2 = \ln\left(\frac{em}{2s_2}\right)\alpha$, $\alpha_1 = \alpha$, one has $C_4\delta^2 - \alpha_1 - \alpha_2 > 0$ as long as $\frac{s_2}{m} = \alpha$ be small enough and therefore the probability $\exp(-(C_4\delta^2 - \alpha_1 - \alpha_2)m)$ be exponentially small.

**Lemma C.1.3** (bounding $\|w^{(2)}\|_2$) if $c_1 > \frac{94}{\tilde{c}}$ and conditions in lemma C.1.2 hold, then one has,

$$prob\left(\|w^{(2)}\|_2 < \frac{1}{4}\right) \geq (1-\varepsilon)\left(1-\frac{\varepsilon}{n}\right)$$

Proof:

By definition in step 2 of golfing scheme,

$$w^{(2)} = \sigma_x - u^{(2)}(s_x) = w^{(1)} - \frac{1}{|\Lambda_1|}A^T(s_x,\Lambda_1)A(\Lambda_1,s_x)w^{(1)} \qquad (C.1.3)$$

The second equality of (C.1.3) holds because,
$w^{(2)} = \sigma_x - u^{(2)}(s_x) = \sigma_x - A_{|\theta}^T h^{(2)} = \sigma_x - A_{|\theta}^T \Delta h^{(1)}(s_x) - A_{|\theta}^T \Delta h^{(2)}(s_x) =$

$w^{(1)} - \frac{2|s_f|}{m-|s_f|}A_{|\theta}^T(s_x,\Lambda_2)A_{|\theta}(\Lambda_2,s_x)w^{(1)}$. According to (A.5) in lemma A.2, letting $t = \frac{\sqrt{2}}{4(1+\sqrt{2})}\frac{1}{\sqrt{s_x}}$

yields,

$$prob\left(\left\|I_{|s_x|} - \frac{1}{|\Lambda_1|}A^T(s_x,\Lambda_1)A(\Lambda_1,s_x)\right\|_{2\to 2} < \frac{\sqrt{2}}{4(1+\sqrt{2})}\frac{1}{\sqrt{s_x}}\right) \geq 1 - 2\exp\left(-\tilde{c}\frac{|\Lambda_1|}{47|s_x|}\right) \geq 1 -$$

$$2\exp\left(-\tilde{c}\frac{c_1\ln(2n/\varepsilon)}{94}\right) \qquad (C.1.4)$$

where the last inequality of (C.1.4) holds because, $|\Lambda_1| = \frac{m-|s_f|}{2} \geq \frac{c_1}{2}|s_x|\ln(2n/\varepsilon)$. Combining

(C.1.3) and (C.1.4) implies that if $c_1 \geq \frac{94}{\tilde{c}}$, then one has,

$$prob\left(\|w^{(2)}\|_2 < \frac{\sqrt{2}}{4(1+\sqrt{2})}\frac{1}{\sqrt{s_x}}\|w^{(1)}\|_2\bigg|\|w^{(1)}\|_2 < (1+1/\sqrt{2})\sqrt{|s_x|}\right) \geq 1 - \frac{\varepsilon}{n} \qquad (C.1.5)$$

and thus,

$$prob\left(\|w^{(2)}\|_2 < \frac{1}{4}\right) \geq$$

$$prob\left(\|w^{(2)}\|_2 < \frac{\sqrt{2}}{4(1+\sqrt{2})}\frac{1}{\sqrt{s_x}}\|w^{(1)}\|_2\bigg|\|w^{(1)}\|_2 < (1+1/\sqrt{2})\sqrt{|s_x|}\right)prob\left(\|w^{(1)}\|_2 < (1+$$

$$1/\sqrt{2})\sqrt{|s_x|}\right) \geq (1-\varepsilon)\left(1-\frac{\varepsilon}{n}\right) \qquad (C.1.6)$$

which proves (C.1.3). ∎

**Lemma C.1.4** (bounding $\|w^{(3)}\|_2$) if $c_1 \geq \frac{8c_I^2|s_f|}{\tilde{c}(m-|s_f|)}$ and conditions in lemma C.1.3 hold, then one has,

$$prob\left(\|w^{(3)}\|_2 \leq \sqrt{\frac{m-|s_f|}{|s_f|}}\frac{1}{8c_I\sqrt{\ln(2n/\varepsilon)}}\right) \geq (1-2exp(-|s_x|))(1-\varepsilon)\left(1-\frac{\varepsilon}{n}\right) \qquad (C.1.7)$$

Proof:

Since by the definition of $w^{(3)}$ in the golfing scheme,

$$w^{(3)} = \sigma_x - u^{(2)}(s_x) = (I_{|s_x|} - \frac{1}{|\Lambda_2|}A^T(s_x,\Lambda_2)A(\Lambda_2,s_x))w^{(2)} \qquad (C.1.8)$$

According to (A.5) in lemma A.2, letting $t = \sqrt{\frac{m-|s_f|}{|s_f|}}\frac{1}{2c_I\sqrt{\ln(2n/\varepsilon)}}$,

$$\text{prob}\left(\left\|I_{|s_x|} - \frac{1}{|\Lambda_2|}A^T(s_x,\Lambda_2)A(\Lambda_2,s_x)\right\|_{2\to 2} \leq \sqrt{\frac{m-|s_f|}{|s_f|}}\frac{1}{2c_I\sqrt{\ln(2n/\varepsilon)}}\right) \geq$$

$$1 - 2\exp\left(-\tilde{c}\frac{(m-|s_f|)^2}{8c_I^2\ln(2n/\varepsilon)|s_f|}\right) \geq 1 - 2\exp\left(-\tilde{c}\frac{c_1|s_x|(m-|s_f|)}{8c_I^2|s_f|}\right) \quad \text{(C.1.9)}$$

(C.1.9) implies that, if $c_1 \geq \frac{8c_I^2|s_f|}{\tilde{c}(m-|s_f|)}$, then,

$$\text{prob}\left(\|w^{(3)}\|_2 \leq \sqrt{\frac{m-|s_f|}{|s_f|}}\frac{\|w^{(2)}\|_2}{2c_I\sqrt{\ln(2n/\varepsilon)}}\right) \geq 1 - 2\exp(-|s_x|) \quad \text{(C.1.10)}$$

Finally, one has,

$$\text{prob}\left(\|w^{(3)}\|_2 \leq \sqrt{\frac{m-|s_f|}{|s_f|}}\frac{1}{8c_I\sqrt{\ln(2n/\varepsilon)}}\right) \geq$$

$$\text{prob}\left(\|w^{(3)}\|_2 \leq \sqrt{\frac{m-|s_f|}{|s_f|}}\frac{\|w^{(2)}\|_2}{2c_I\sqrt{\ln(2n/\varepsilon)}}\bigg|\|w^{(2)}\|_2 \leq \tfrac{1}{4}\right)\text{prob}\left(\|w^{(2)}\|_2 \leq \tfrac{1}{4}\right) \geq$$

$$(1 - 2\exp(-|s_x|))(1-\varepsilon)\left(1 - \frac{\varepsilon}{n}\right) \quad \text{(C.1.11)}$$

where the last inequality of (C.1.11) follows from (C.1.10) and (C.1.3).∎

## C.2 bounding $\|\Delta u^{(i)}(s_x^c)\|_\infty$

**Lemma C.2.1** (bounding $\|\Delta u^{(1)}(s_x^c)\|_\infty$) if $c_I \geq \sqrt{8c}$, then one has,

$$prob\left(\|\Delta u^{(1)}(s_x^c)\|_\infty < \frac{1}{\sqrt{2}}\right) \geq 1 - \varepsilon = P_{\Delta u}^{(1)} \quad \text{(C.2.1)}$$

Proof: this follows from (C.1.1) in lemma C.1.1.∎

**Lemma C.2.2** (bounding $\|\Delta u^{(2)}(s_x^c)\|_\infty$) if $c_1 \geq 3672c$ and conditions in lemma C.1.2 hold, then one has,

$$prob\left(|\Delta u^{(2)}(i)| < \frac{\sqrt{2}-1}{3\sqrt{2}}, i \in s_x^c\right) \geq (1-\varepsilon)\left(1 - 2exp\left(-\frac{\tilde{c}(m-|s_f|)}{8}\right) - \varepsilon\right) = P_{\Delta u}^{(2)} \quad \text{(C.2.2)}$$

Proof:

Firstly, $\Delta h^{(2)} = \frac{2|s_f|}{m-|s_f|}A_{|\theta}(\Lambda_1, s_x)w^{(1)}$, by letting $t = \frac{1}{2}$ in (A.5) of lemma A.2, one has,

$$\text{prob}\left(\left\|\sqrt{\frac{2|s_f|}{m-|s_f|}}A_{|\theta}(\Lambda_1, s_x)\right\|_{2\to 2} \leq \sqrt{\frac{3}{2}}\right) \geq 1 - 2\exp\left(-\frac{\tilde{c}(m-|s_f|)}{8}\right) \quad \text{(C.2.3)}$$

since $c_I \geq \sqrt{8c}$, combining (C.2.3) the fact stated in lemma C.1.2 that $\text{prob}\left(\|w^{(1)}\|_2 < (1 + 1/\sqrt{2})\sqrt{|s_x|}\right) \geq 1 - \varepsilon$ and then applying a union bound yields,

$$\text{prob}\left(\|\Delta h^{(2)}\|_2 \leq \sqrt{\frac{2|s_f|}{m-|s_f|}}\sqrt{\frac{3}{2}}(1+1/\sqrt{2})\sqrt{|s_x|}\right) \geq 1 - 2\exp\left(-\frac{\tilde{c}(m-|s_f|)}{8}\right) - \varepsilon \quad \text{(C.2.4)}$$

Secondly, since $\Delta u^{(2)}(i) = \langle A(\Lambda_1, i), \theta_A \Delta h^{(2)}\rangle, i \in s_x^c$, letting $t = \frac{\sqrt{2}-1}{3\sqrt{2}}$ in (A.4) of lemma A.2 yields,

$$\text{prob}\left(|\Delta u^{(2)}(i)| < \frac{\sqrt{2}-1}{3\sqrt{2}}\right) \geq 1 - 2\exp\left(-\frac{1}{420c\theta_A^2 \|\Delta h^{(2)}\|_2^2}\right) \quad (C.2.5)$$

because,

$$\text{prob}\left(|\Delta u^{(2)}(i)| < \frac{\sqrt{2}-1}{3\sqrt{2}}\right) \geq$$

$$\text{prob}\left(|\Delta u^{(2)}(i)| < \frac{\sqrt{2}-1}{3\sqrt{2}} \middle| \|\Delta h^{(2)}\|_2 \leq \sqrt{\frac{2|s_f|}{m-|s_f|}}\sqrt{\frac{3}{2}}(1+1/\sqrt{2})\sqrt{|s_x|}\right) \text{prob}\left(\|\Delta h^{(2)}\|_2 \leq \sqrt{\frac{2|s_f|}{m-|s_f|}}\sqrt{\frac{3}{2}}(1+1/\sqrt{2})\sqrt{|s_x|}\right) \quad (C.2.6)$$

and according to (C.2.5),

$$\text{prob}\left(|\Delta u^{(2)}(i)| < \frac{\sqrt{2}-1}{3\sqrt{2}} \middle| \|\Delta h^{(2)}\|_2 \leq \sqrt{\frac{2|s_f|}{m-|s_f|}}\sqrt{\frac{3}{2}}(1+1/\sqrt{2})\sqrt{|s_x|}\right) \geq 1 - 2\exp\left(-\frac{m-|s_f|}{3672c|s_x|}\right) \quad (C.2.7)$$

since $m - |s_f| \geq c_1|s_x|\ln(2n/\varepsilon)$ which implies that if $c_1 \geq 3672c$, then the probability in the left-most side of (C.2.7) is larger than $1 - \frac{\varepsilon}{n}$, applying a union bound, one has,

$$\text{prob}\left(|\Delta u^{(2)}(i)| < \frac{\sqrt{2}-1}{3\sqrt{2}}, i \in s_x^c \middle| \|\Delta h^{(2)}\|_2 \leq \sqrt{\frac{2|s_f|}{m-|s_f|}}\sqrt{\frac{3}{2}}(1+1/\sqrt{2})\sqrt{|s_x|}\right) \geq 1 - \varepsilon \quad (C.2.8)$$

combining (C.2.6), (C.2.8) and (C.2.4) yields,

$$\text{prob}\left(|\Delta u^{(2)}(i)| < \frac{\sqrt{2}-1}{3\sqrt{2}}, i \in s_x^c\right) \geq (1-\varepsilon)\left(1 - 2\exp\left(-\frac{\tilde{c}(m-|s_f|)}{8}\right) - \varepsilon\right) \quad (C.2.9)$$

which proves the lemma. ∎

**Lemma C.2.3** (bounding $\|\Delta u^{(3)}(s_x^c)\|_\infty$) if $c_1|s_x| \geq 79c$ and conditions in lemma C.1.3 hold, then one has,

$$\text{prob}\left(\|\Delta u^{(3)}(s_x^c)\|_\infty < \frac{\sqrt{2}-1}{3\sqrt{2}}\right) \geq (1-\varepsilon)\left(1 - 2\exp\left(-\frac{\tilde{c}(m-|s_f|)}{8}\right) - \varepsilon - \frac{\varepsilon}{n}\right) = P_{\Delta u}^{(3)} \quad (C.2.10)$$

Proof:

We follow the argument similarly as in lemma C.2.2, firstly, by definition in the golfing scheme, $\Delta h^{(3)} = \frac{2|s_f|}{m-|s_f|}A_{|\theta}(\Lambda_2, s_x)w^{(2)}$, by letting $t = \frac{1}{2}$ in (A.5) of lemma A.2, one has,

$$\text{prob}\left(\left\|\sqrt{\frac{2|s_f|}{m-|s_f|}}A_{|\theta}(\Lambda_2, s_x)\right\|_{2\to 2} \leq \sqrt{\frac{3}{2}}\right) \geq 1 - 2\exp\left(-\frac{\tilde{c}(m-|s_f|)}{8}\right) \quad (C.2.11)$$

combining the fact stated in lemma C.1.3 that $\text{prob}\left(\|w^{(2)}\|_2 < \frac{1}{4}\right) \geq (1-\varepsilon)\left(1 - \frac{\varepsilon}{n}\right)$ and then applying a union bound yields,

$$\text{prob}\left(\|\Delta h^{(3)}\|_2 \leq \sqrt{\frac{2|s_f|}{m-|s_f|}}\frac{1}{4}\sqrt{\frac{3}{2}}\right) \geq 1 - 2\exp\left(-\frac{\tilde{c}(m-|s_f|)}{8}\right) - \varepsilon - \frac{\varepsilon}{n} \quad (C.2.12)$$

Secondly, since $\Delta u^{(3)}(i) = \langle A(\Lambda_2, i), \theta_A \Delta h^{(3)}\rangle, i \in s_x^c$, letting $t = \frac{\sqrt{2}-1}{3\sqrt{2}}$ in (A.4) of lemma A.2 yields,

$$\text{prob}\left(|\Delta u^{(3)}(i)| < \frac{\sqrt{2}-1}{3\sqrt{2}}\right) \geq 1 - 2\exp\left(-\frac{1}{420c\theta_A^2 \|\Delta h^{(3)}\|_2^2}\right) \quad (C.2.13)$$

because,

$$\text{prob}\left(|\Delta u^{(3)}(i)| < \frac{\sqrt{2}-1}{3\sqrt{2}}\right) \geq$$

$$\text{prob}\left(|\Delta u^{(3)}(i)| < \frac{\sqrt{2}-1}{3\sqrt{2}}\middle| \|\Delta h^{(3)}\|_2 \leq \frac{1}{4}\sqrt{\frac{3}{2}}\sqrt{\frac{2|s_f|}{m-|s_f|}}\right)\text{prob}\left(\|\Delta h^{(3)}\|_2 \leq \frac{1}{4}\sqrt{\frac{3}{2}}\sqrt{\frac{2|s_f|}{m-|s_f|}}\right)$$

(C.2.14)

and according to (C.2.13)

$$\text{prob}\left(|\Delta u^{(3)}(i)| < \frac{\sqrt{2}-1}{3\sqrt{2}}\middle| \|\Delta h^{(3)}\|_2 \leq \frac{1}{4}\sqrt{\frac{3}{2}}\sqrt{\frac{4|s_f|}{m-|s_f|}}\right) \geq 1 - 2\exp\left(-\frac{m-|s_f|}{79c}\right) \geq$$

$$1 - 2\exp\left(-\frac{c_1|s_x|\ln(2n/\varepsilon)}{79c}\right) \quad \text{(C.2.15)}$$

(C.2.15) implies that if $c_1|s_x| \geq 79c$, then, the left most side of (C.2.15) is larger than $1 - \frac{\varepsilon}{n}$,

combining (C.2.14), (C.2.15) and (C.2.12) and then apply a union bound argument yields,

$$\text{prob}\left(|\Delta u^{(3)}(i)| < \frac{\sqrt{2}-1}{3\sqrt{2}}, i \in s_x^c\right) \geq (1-\varepsilon)\left(1 - 2\exp\left(-\frac{\tilde{c}(m-|s_f|)}{8}\right) - \varepsilon - \frac{\varepsilon}{n}\right) \quad \text{(C.2.16)}$$

which proves the lemma.■

**lemma C.2.4** (bounding $\|\Delta u^{(4)}(s_x^c)\|_\infty$) if $|s_f|c_I^2 \geq 40c$ and conditions in lemma C.1.4 hold, then one has,

$$prob\left(\|\Delta u^{(4)}(s_x^c)\|_\infty < \frac{\sqrt{2}-1}{3\sqrt{2}}\right) \geq (1-\varepsilon)\left(1 - 2exp\left(-\frac{\tilde{c}(m-|s_f|)}{4}\right) - 2exp(-|s_x|) - \varepsilon - \frac{\varepsilon}{n}\right) =$$

$$P_{\Delta u}^{(4)} \quad \text{(C.2.17)}$$

Proof:

We follow the argument similarly as in lemma C.2.2, firstly, by definition in the golfing scheme,

$\Delta h^{(4)} == A_{|\theta}^\dagger(\Lambda_3, s_x)w^{(3)}$, by letting $t = \frac{1}{2}$ in (2.1.6) of lemma 2.1.4, one has,

$$\text{prob}\left(\|A_{|\theta}^\dagger(\Lambda_3, s_x)\|_{2\to 2} \leq 2\sqrt{\frac{3}{2}}\sqrt{\frac{|s_f|}{m-|s_f|}}\right) \geq 1 - 2\exp\left(-\frac{\tilde{c}(m-|s_f|)}{4}\right) \quad \text{(C.2.18)}$$

combining the fact stated in lemma C.1.4 that $\text{prob}\left(\|w^{(3)}\|_2 \leq \sqrt{\frac{m-|s_f|}{|s_f|}}\frac{1}{8c_I\sqrt{\ln(2n/\varepsilon)}}\right) \geq$

$(1 - 2\exp(-|s_x|))(1-\varepsilon)\left(1 - \frac{\varepsilon}{n}\right)$ and then applying a union bound yields,

$$\text{prob}\left(\|\Delta h^{(4)}\|_2 \leq \sqrt{6}\frac{1}{8c_I\sqrt{\ln(2n/\varepsilon)}}\right) \geq 1 - 2\exp\left(-\frac{\tilde{c}(m-|s_f|)}{4}\right) - 2\exp(-|s_x|) - \varepsilon - \frac{\varepsilon}{n}$$

(C.2.19)

Secondly, since $\Delta u^{(4)}(i) = \langle A(\Lambda_3, i), \theta_A \Delta h^{(4)}\rangle, i \in s_x^c$, letting $t = \frac{\sqrt{2}-1}{3\sqrt{2}}$ in (A.4) of lemma A.1 yields,

$$\text{prob}\left(|\Delta u^{(4)}(i)| < \frac{\sqrt{2}-1}{3\sqrt{2}}\right) \geq 1 - 2\exp\left(-\frac{1}{420c\theta_A^2\|\Delta h^{(4)}\|_2^2}\right) \quad \text{(C.2.20)}$$

because,

$$\text{prob}\left(|\Delta u^{(4)}(i)| < \tfrac{\sqrt{2}-1}{3\sqrt{2}}\right) \geq$$

$$\text{prob}\left(|\Delta u^{(4)}(i)| < \tfrac{\sqrt{2}-1}{3\sqrt{2}} \mid \|\Delta h^{(4)}\|_2 \leq \sqrt{6}\tfrac{1}{8c_I\sqrt{\ln(2n/\varepsilon)}}\right) \text{prob}\left(\|\Delta h^{(4)}\|_2 \leq \sqrt{6}\tfrac{1}{8c_I\sqrt{\ln(2n/\varepsilon)}}\right)$$

(C.2.21)

and according to (C.2.20)

$$\text{prob}\left(|\Delta u^{(4)}(i)| < \tfrac{\sqrt{2}-1}{3\sqrt{2}} \mid \|\Delta h^{(4)}\|_2 \leq \sqrt{6}\tfrac{1}{8c_I\sqrt{\ln(2n/\varepsilon)}}\right) \geq 1 - 2\exp\left(-\tfrac{64|s_f|c_I^2}{420 \times 6c}\ln(2n/\varepsilon)\right)$$

(C.2.22)

(C.2.22) implies that if $|s_f|c_I^2 \geq 40c$, then the left most side of (C.2.22) is larger than $1 - \tfrac{\varepsilon}{n}$, combining (C.2.21), (C.2.22) and (C.2.10) and then apply a union bound argument yields,

$$\text{prob}\left(|\Delta u^{(4)}(i)| < \tfrac{\sqrt{2}-1}{3\sqrt{2}}, i \in s_x^c\right) \geq (1-\varepsilon)\left(1 - 2\exp\left(-\tfrac{\tilde{c}(m-|s_f|)}{4}\right) - 2\exp(-|s_x|) - \varepsilon - \tfrac{\varepsilon}{n}\right)$$

(C.2.23)

which proves the lemma. ∎

## C.3 bounding $\|\Delta h^{(i)}(s_e^c)\|_\infty$

**Lemma C.3.1** (bounding $\|\Delta h^{(2)}(s_f^c)\|_\infty$) if $c_1 \geq 158 c_I^2 \tfrac{|s_f|}{m-|s_f|}$ and conditions in lemma C.1.2 hold, then one has,

$$prob\left(\|\Delta h^{(2)}(s_f^c)\|_\infty < \tfrac{1}{3\theta_I}\right) \geq 1 - 2\exp\left(-\tfrac{\tilde{c}(m-|s_f|)}{8}\right) - \varepsilon = P_{\Delta h}^{(2)} \qquad (C.3.1)$$

Proof:

Suppose the inverse event $\|\Delta h^{(2)}(s_f^c)\|_\infty \geq \tfrac{1}{3\theta_I}$ is true, let $\Lambda = s_f^c \backslash \Lambda_1$, $q = \sqrt{\tfrac{2|s_f|}{m-|s_f|}} A_{|\theta}(\Lambda, s_x) w^{(1)}$, by the definition of $\Lambda_1$ in the golfing scheme, the inverse event implies that $\|q\|_2 \geq \tfrac{1}{3\theta_I}\sqrt{\tfrac{m-|s_f|}{2|s_f|}}\sqrt{\tfrac{m-|s_f|}{2}}$, therefore,

$$prob\left(\|\Delta h^{(2)}(s_f^c)\|_\infty \geq \tfrac{1}{3\theta_I}\right) \leq prob\left(\|q\|_2 \geq \tfrac{1}{3\theta_I}\sqrt{\tfrac{m-|s_f|}{2|s_f|}}\sqrt{\tfrac{m-|s_f|}{2}}\right) \qquad (C.3.2)$$

to find an upper bound for the right hand side of (C.3.2), notice that,

$$prob\left(\left\|\sqrt{\tfrac{2|s_f|}{m-|s_f|}} A_{|\theta}(\Lambda, s_x)\right\|_{2 \to 2} \leq \sqrt{\tfrac{3}{2}}\right) \geq 1 - 2\exp\left(-\tfrac{\tilde{c}(m-|s_f|)}{8}\right) \qquad (C.3.3)$$

where (C.3.3) follows from (A.5) in lemma A.2 by letting t=1/2, combining (C.3.3) with the fact stated in lemma C.1.2 that $prob\left(\|w^{(1)}\|_2 < (1+1/\sqrt{2})\sqrt{|s_x|}\right) \geq 1-\varepsilon$, and then applying a union bound yields,

$$prob\left(\|q\|_2 < \sqrt{\tfrac{3}{2}}(1+1/\sqrt{2})\sqrt{|s_x|}\right) \geq 1 - 2\exp\left(-\tfrac{\tilde{c}(m-|s_f|)}{8}\right) - \varepsilon \qquad (C.3.4)$$

(C.3.4) implies,

$$prob\left(\|q\|_2 \geq \sqrt{\tfrac{3}{2}}(1+1/\sqrt{2})\sqrt{|s_x|}\right) < 2\exp\left(-\tfrac{\tilde{c}(m-|s_f|)}{8}\right) + \varepsilon \qquad (C.3.5)$$

according to (C.3.5), if below (C.3.6) holds,

$$\frac{1}{3\theta_I}\sqrt{\frac{m-|s_f|}{2|s_f|}}\sqrt{\frac{m-|s_f|}{2}} \geq \sqrt{\frac{3}{2}}(1+1/\sqrt{2})\sqrt{|s_x|} \qquad (C.3.6)$$

Then,

$$\text{prob}\left(\|q\|_2 \geq \frac{1}{3\theta_I}\sqrt{\frac{m-|s_f|}{2}}\right) \leq \text{prob}\left(\|q\|_2 \geq \sqrt{\frac{3}{2}\frac{5}{4}}\sqrt{|s_x|}\right) < 2exp\left(-\frac{\tilde{c}(m-|s_f|)}{8}\right) + \varepsilon \qquad (C.3.7)$$

using the assumption $m - |s_f| \geq c_1|s_x|\ln(2n/\varepsilon)$, a sufficient condition for (C.3.6) is,

$$c_1 \geq 158c_I^2 \frac{|s_f|}{m-|s_f|} \qquad (C.3.8)$$

Therefore, if (C.3.8) holds, then (C.3.7) holds, using (C.3.2), one may conclude (C.3.1) holds, which proves the lemma.∎

**Lemma C.3.2** (bounding $\|\Delta h^{(3)}(s_f^c)\|_\infty$) if $m - |s_f| \geq 3.4c_I^2 \frac{|s_f|}{m-|s_f|}\ln(2n/\varepsilon)$ and conditions in lemma C.1.3 hold, then one has,

$$\text{prob}\left(\|\Delta h^{(3)}(s_f^c)\|_\infty < \frac{1}{3\theta_I}\right) \geq 1 - 2exp\left(-\frac{\tilde{c}(m-|s_f|)}{8}\right) - \frac{(n+1)\varepsilon}{n} = P_{\Delta h}^{(3)} \qquad (C.3.9)$$

Proof:

The proof follows the arguments from the previous lemma C.3.1, suppose the inverse event $\|\Delta h^{(3)}(s_f^c)\|_\infty \geq \frac{1}{3\theta_I}$ is true, let $\wedge = s_f^c \backslash \wedge_2$, $q = \sqrt{\frac{2|s_f|}{m-|s_f|}} A_{|\theta}(\wedge, s_x)w^{(2)}$, by the definition of $\wedge_2$ in the golfing scheme, the inverse event implies that $\|q\|_2 \geq \frac{1}{3\theta_I}\sqrt{\frac{m-|s_f|}{2|s_f|}}\sqrt{\frac{m-|s_f|}{2}}$, therefore,

$$\text{prob}\left(\|\Delta h^{(2)}(s_f^c)\|_\infty \geq \frac{1}{3\theta_I}\right) \leq \text{prob}\left(\|q\|_2 \geq \frac{1}{3\theta_I}\sqrt{\frac{m-|s_f|}{2|s_f|}}\sqrt{\frac{m-|s_f|}{2}}\right) \qquad (C.3.10)$$

to find an upper bound for the right hand side of (C.3.10), notice that,

$$\text{prob}\left(\sqrt{\frac{2|s_f|}{m-|s_f|}}\|A_{|\theta}(\wedge, s_x)\|_{2\to 2} \leq \sqrt{\frac{3}{2}}\right) \geq 1 - 2exp\left(-\frac{\tilde{c}(m-|s_f|)}{8}\right) \qquad (C.3.11)$$

where (C.3.11) follows from (A.5) in lemma A.2, combining (C.3.11) with the fact stated in lemma (C.1.3) that $\text{prob}\left(\|w^{(2)}\|_2 < \frac{1}{4}\right) \geq (1-\varepsilon)\left(1-\frac{\varepsilon}{n}\right)$, applying a union bound yields,

$$\text{prob}\left(\|q\|_2 < \sqrt{\frac{3}{2}\frac{1}{4}}\right) \geq 1 - 2exp\left(-\frac{\tilde{c}(m-|s_f|)}{8}\right) - \frac{(n+1)\varepsilon}{n} \qquad (C.3.12)$$

(C.3.12) implies,

$$\text{prob}\left(\|q\|_2 \geq \sqrt{\frac{3}{2}\frac{1}{4}}\right) < 2exp\left(-\frac{\tilde{c}(m-|s_f|)}{8}\right) + \frac{(n+1)\varepsilon}{n} \qquad (C.3.13)$$

according to (C.3.13), if below (C.3.14) holds,

$$\frac{1}{3\theta_I}\sqrt{\frac{m-|s_f|}{2|s_f|}}\sqrt{\frac{m-|s_f|}{2}} \geq \sqrt{\frac{3}{2}\frac{1}{4}} \qquad (C.3.14)$$

Then,

$$\text{prob}\left(\|q\|_2 \geq \frac{1}{3\theta_I}\sqrt{\frac{m-|s_f|}{2}}\right) \leq \text{prob}\left(\|q\|_2 \geq \sqrt{\frac{3}{2}\frac{1}{4}}\right) < 2exp\left(-\frac{\tilde{c}(m-|s_f|)}{8}\right) + \frac{(n+1)\varepsilon}{n} \qquad (C.3.15)$$

a sufficient condition for (C.3.14) is,

$$m - |s_f| \geq 3.4c_I^2 \frac{|s_f|}{m-|s_f|}\ln(2n/\varepsilon) \qquad (C.3.16)$$

Therefore, if (C.3.16) holds, then (C.3.15) holds, using (C.3.10), one may conclude (C.3.9) holds, which proves the lemma.∎

**Lemma C.3.3** (bounding $\|\Delta h^{(4)}(s_f^c)\|_\infty$) if conditions in lemma C.1.4 holds, then one has,

$$prob\left(\|\Delta h^{(4)}(s_f^c)\|_\infty < \frac{1}{3\theta_I}\right) \geq 1 - 2\exp\left(-\frac{\tilde{c}(m-|s_f|)}{4}\right) - 2\exp(-|s_x|) - \frac{(n+1)\varepsilon}{n} = P_{\Delta h}^{(4)}$$

*(C.3.17)*

Proof:

By the definition of $\Delta h^{(4)}$, one has,

$$\|\Delta h^{(4)}\|_2 \leq \|A_{|\theta}^\dagger(\Lambda_3, s_x)\|_{2\to 2} \|w^{(3)}\|_2 \quad (C.3.17)$$

Notice that,

$$prob\left(\|A_{|\theta}^\dagger(\Lambda_3, s_x)\|_{2\to 2} < \sqrt{6}\sqrt{\frac{|s_f|}{m-|s_f|}}\right) \geq 1 - 2\exp\left(-\frac{\tilde{c}(m-|s_f|)}{4}\right) \quad (C.3.18)$$

where (C.3.18) follows from (A.5) in lemma A.3 by letting $t = \frac{1}{2}$, combining (C.3.18) and the fact stated in (C.1.7) that $prob\left(\|w^{(3)}\|_2 \leq \sqrt{\frac{m-|s_f|}{|s_f|}} \frac{1}{8c_I\sqrt{\ln(2n/\varepsilon)}}\right) \geq (1 - 2\exp(-|s_x|))(1-\varepsilon)\left(1 - \frac{\varepsilon}{n}\right)$, applying a union bound yields,

$$prob\left(\|\Delta h^{(4)}\|_2 < \frac{1}{3} \frac{1}{c_I\sqrt{\ln(2n/\varepsilon)}}\right) \geq 1 - 2\exp\left(-\frac{\tilde{c}(m-|s_f|)}{4}\right) - 2\exp(-|s_x|) - \frac{(n+1)\varepsilon}{n} \quad (C.3.19)$$

which proves the conclusion of the lemma.∎

## C.4. Proving that B in lemma 3.2 is a full rank matrix

**Theorem C.4.1** if $|s_f^c| \geq \frac{8}{3\tilde{c}}(7|s_x| + 2\ln(2\varepsilon^{-1}))$, then with probability at least $1-\varepsilon$, where $\varepsilon \in (0,1)$ is a constant, matrix $B = [\theta_A A([m], s_x), \theta_I I_m([m], s_f)]$ is full rank matrix.

Proof: since $\theta_A, \theta_I > 0$, let $\lambda = \frac{\theta_A}{\theta_I} > 0$, to show B is a full rank matrix, it's sufficient to show that $\tilde{B} = [\lambda A([m], s_x), I_m([m], s_f)]$ is a full rank matrix, to this end, it's sufficient to show that matrix $C = \tilde{B}^T \tilde{B}$ is non-singular, Since,

$$C = \begin{bmatrix} \lambda^2 A^T(s_x,[m])A([m],s_x) & \lambda A^T(s_x,s_f) \\ \lambda A(s_f,s_x) & I_{|s_f|} \end{bmatrix} \quad (C.4.1)$$

By Schur complement decomposition, one has,

$$HCH^T = \begin{bmatrix} \lambda^2 A^T(s_x,s_f^c)A(s_f^c,s_x) & 0 \\ 0 & I_{|s_f|} \end{bmatrix} \quad (C.4.2)$$

Where $H = \begin{bmatrix} I_{|s_x|} & -\lambda A^T(s_x,s_f) \\ 0 & I_{|s_f|} \end{bmatrix}$ is a non-singular matrix, then letting $\delta = 0.5$ in lemma A.3, it follows that $\left\|\frac{1}{m}\lambda^2 A^T(s_x,[m])A([m],s_x)\right\|_{2\to 2} \geq \frac{\lambda^2}{2}$ holds with probability at least $1-\varepsilon$, if

$|s_f^c| \geq \frac{8}{3\tilde{c}}(7|s_x| + 2\ln(2\varepsilon^{-1}))$ holds, which proves the conclusion of the theorem. ∎

## C.5. proof of theorem 2.2

Proof:

The conclusion follows by combining results from C.1~C.4 together and then applying a union bound.∎